\documentclass[%
 reprint,
superscriptaddress,
 amsmath,amssymb,
prmaterials,
]{revtex4-2}

\usepackage{graphicx}
\usepackage{dcolumn}
\usepackage{bm}

\begin{document}

\preprint{APS/123-QED}

\title{Growth and characterization of $\alpha$-Sn thin films on In- and Sb-rich reconstructions of InSb(001)}

\author{Aaron N. Engel}
\email{aengel@ucsb.edu}
\affiliation{Materials Department, University of California Santa Barbara, Santa Barbara, California 93106, USA}

\author{Connor P. Dempsey}
\affiliation{%
Electrical and Computer Engineering Department, University of California Santa Barbara, Santa Barbara, California 93106, USA}%

\author{Hadass S. Inbar}
\affiliation{Materials Department, University of California Santa Barbara, Santa Barbara, California 93106, USA}

\author{Jason T. Dong}
\affiliation{Materials Department, University of California Santa Barbara, Santa Barbara, California 93106, USA}

\author{Shinichi Nishihaya}
\altaffiliation[]{Present Address: Department of Physics, Tokyo Institute of Technology, Tokyo 152-8551, Japan}
\affiliation{%
 Electrical and Computer Engineering Department, University of California Santa Barbara, Santa Barbara, California 93106, USA}%

\author{Yuhao Chang}
\affiliation{Materials Department, University of California Santa Barbara, Santa Barbara, California 93106, USA}

\author{Alexei V. Fedorov}
\affiliation{
Advanced Light Source, Lawrence Berkeley National Laboratory, Berkeley, California 94720, USA
}%
\author{Makoto Hashimoto}
\affiliation{
 Stanford Synchrotron Radiation Lightsource, SLAC National Accelerator Laboratory, 2575 Sand Hill Road, Menlo Park, California 94025, USA
}%
\author{Donghui Lu}
\affiliation{
 Stanford Synchrotron Radiation Lightsource, SLAC National Accelerator Laboratory, 2575 Sand Hill Road, Menlo Park, California 94025, USA
}%

\author{Christopher J. Palmstr\o{}m}
\email{cjpalm@ucsb.edu}
\affiliation{Materials Department, University of California Santa Barbara, Santa Barbara, California 93106, USA}
\affiliation{%
 Electrical and Computer Engineering Department, University of California Santa Barbara, Santa Barbara, California 93106, USA}%

\date{\today}

\begin{abstract}
$\alpha$-Sn thin films can exhibit a variety of topologically non-trivial phases. Both studying the transitions between these phases and making use of these phases in eventual applications requires good control over the electronic and structural quality of $\alpha$-Sn thin films. $\alpha$-Sn growth on InSb often results in out-diffusion of indium, a \textit{p}-type dopant. By growing $\alpha$-Sn via molecular beam epitaxy on the Sb-rich c(4$\times$4) surface reconstruction of InSb(001) rather than the In-rich c(8$\times$2), we demonstrate a route to substantially decrease and minimize this indium incorporation. The reduction in indium concentration allows for the study of the surface and bulk Dirac nodes in $\alpha$-Sn via angle-resolved photoelectron spectroscopy without the common approaches of bulk doping or surface dosing, simplifying topological phase identification. The lack of indium incorporation is verified in angle-resolved and -integrated ultraviolet photoelectron spectroscopy as well as in clear changes in the Hall response. 
\end{abstract}

\maketitle


\section{\label{sec:level1}Introduction}
$\alpha$-Sn, the diamond structure allotrope of Sn, is a zero-gap semiconductor with an inverted electronic band structure at the \bm{$\Gamma$} point\cite{Groves1963}. However, the diamond structure is not stable above 286 K in the bulk, transitioning to a trivial metallic tetragonal phase ($\beta$-Sn)\cite{TinPest}. Fortunately, this phase transition temperature can be raised by epitaxial stabilization of $\alpha$-Sn ($a$=6.489 \AA) thin films on closely lattice matched substrates like InSb ($a$=6.479 \AA) and CdTe ($a$=6.482 \AA)\cite{Farrow1981}. Theoretical predictions show that the zero-gap semiconductor phase transitions to a Dirac semimetal (DSM) under epitaxial compressive strain and a topological insulator (TI) under epitaxial tensile strain\cite{Huang2017, Kufner2014}. Experiments have suggested that on InSb, $\alpha$-Sn thin films go through multiple topological phase transitions between 2D TI, 3D TI, 2D DSM, 3D DSM, and normal insulator as a function of surface orientation, strain, and film thickness, generating much interest as a testbed for topological phase transitions\cite{Anh2021,Barbedienne1954,Barfuss2013,Chen2022,Ohtsubo2013,deCoster2018,Rogalev2017,Rogalev2019,Scholz2018,Xu2017}. Exact results vary by research group and measurement technique. $\alpha$-Sn is--to our knowledge--the only elemental material showing many of these phases. Due to its elemental nature, $\alpha$-Sn should not suffer from the well-known point defect issues seen in many other multi-component topological materials. This makes the material very attractive for high-mobility Dirac fermion transport studies. 

Bulk single crystals of $\alpha$-Sn have been well studied in the past, where dopant concentration could be better controlled and the inverted band structure was confirmed\cite{Ewald1968}. In thin film growth of $\alpha$-Sn, InSb has been the primary substrate of choice for surface science studies. These growths of $\alpha$-Sn are usually performed via molecular beam epitaxy (MBE) on sputter-anneal cleaned InSb leading to an In-rich surface. Indium is known to be a \textit{p}-type dopant in $\alpha$-Sn in the bulk\cite{Wagner1971} and has been seen to readily incorporate into $\alpha$-Sn thin films from the InSb substrate\cite{Madarevic2020,Magnano2002}. This effect even occurs when $\alpha$-Sn is grown on a fresh \textit{in-situ} MBE-grown InSb buffer layer with an In-rich surface reconstruction\cite{Chen2022}. The heavy \textit{p}-type doping results in a Fermi level below the valence band maximum (VBM) and the surface Dirac point (near the VBM), making identification of the topological phase difficult via techniques such as angle-resolved photoemission spectroscopy (ARPES).  In addition, the bulk mobility \cite{Wagner1971} and surface state mobility are both inversely dependent on carrier density\cite{Hofmann2022}, furthering the need for reduced unintentional \textit{p}-type doping.

\begin{figure*}
\includegraphics{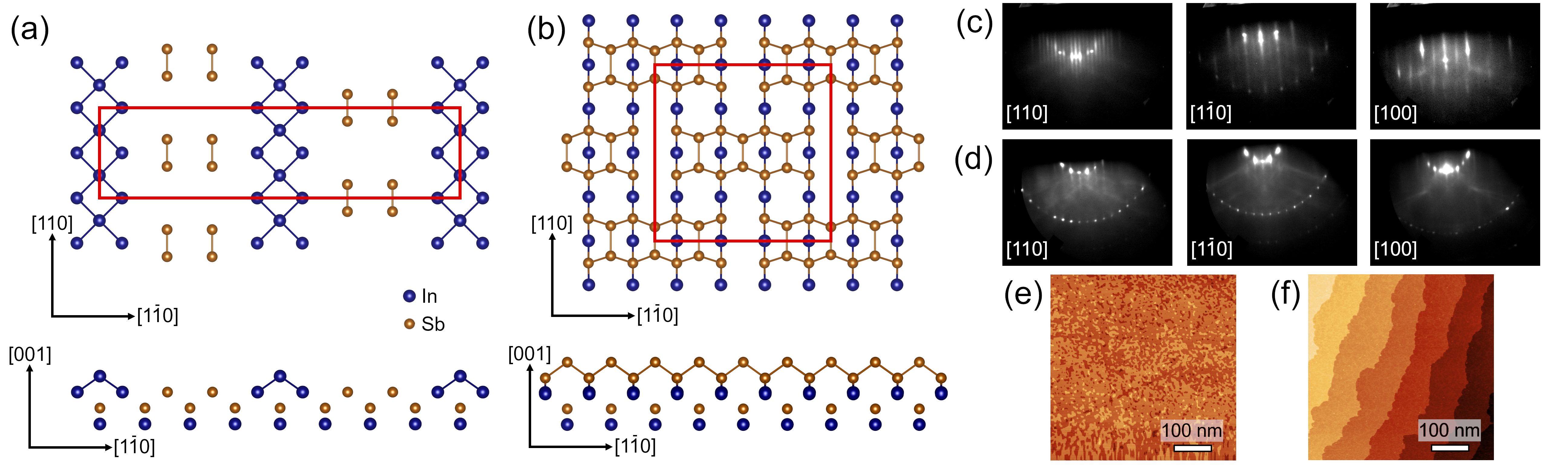}
\caption{\label{Fig1}Surface reconstructions of InSb(001) surface. (a) Top and side view of the In-rich c(8$\times$2) reconstruction following the model of \cite{Davis1999}. (b) Top and side view of the Sb-Rich c(4$\times$4) reconstruction using the atomic positions determined by \cite{Jones1998}. RHEED patterns along the [110],[1$\bar{1}$0], and [100] directions for (c) the c(8$\times$2) reconstruction with a 4$\times$, 2$\times$, and 2$\times$ pattern, respectively, and (d) the c(4$\times$4) reconstruction with a 2$\times$, 2$\times$, and 4$\times$ pattern, respectively. \textit{In-situ} STM images of (c) the c(8$\times$2) surface directly after atomic hydrogen cleaning and (d) the c(4$\times$4) surface after the Sb termination procedure.}
\end{figure*}

ARPES has been a key technique for fingerprinting topological materials due to the direct measurement of the filled-state band structure\cite{Hasan2010,Lv2019}. Fingerprinting via another common technique, magnetotransport, either requires extensive fitting of quantum oscillations via Lifshitz-Kosevich analysis to give indirect evidence of the topological phase or a gated device to map out a relevant portion of the Landau fan diagram for a more direct measurement of the topological phase\cite{Ando2013}. In both cases, the filled bands must have high enough mobilities such that the cyclotron orbit time is shorter than the scattering time in the material\cite{Shoenberg1984}. For $\alpha$-Sn, this condition is frequently not met: only the surface state oscillation is observed in low mobility films\cite{Barbedienne1954,Madarevic2022}, while high mobility films show two oscillation frequencies\cite{Anh2021}. It is difficult to correctly identify the topological phase without measuring both the bulk and surface bands. The two key parameters to differentiate the possible topological phases are the presence (or absence) of a gap in the topological surface states and the presence (or absence) of a gap between the bulk conduction and valence bands. Finally, the presence of parallel bulk transport in indium doped $\alpha$-Sn (under conditions where the topological phase includes a bulk bandgap) greatly reduces the applicability of the material both in fundamental physics studies and devices making use of the topological surface states.

Besides doping, incorporation of indium has been proposed to reduce the quality of $\alpha$-Sn growth\cite{Ohtsubo2013}, similar to what has been suggested in $\alpha$-Ge$_{1-x}$Sn$_x$ growth\cite{Giunto2022}. To reduce \textit{p}-type doping and produce higher quality films, bismuth and tellurium are frequently used as compensatory dopants and/or surfactants\cite{Ohtsubo2013,Scholz2018}. Bismuth and tellurium also possibly change topological signatures from $\alpha$-Sn, modifying at least the band velocity of $\alpha$-Sn’s linear surface states\cite{Madarevic2020}. The structural and electronic effects of incorporating Bi/Te in the $\alpha$-Sn has not been investigated in detail. Another standard solution to limit indium incorporation is to use a low substrate temperature and high Sn growth rate \cite{Madarevic2020,Chen2022,Magnano2002}. However, a lower substrate temperature during growth was seen to be associated with lower structural quality of the epitaxial $\alpha$-Sn\cite{Betti2002}.

Most reported growths studying the topologically non-trivial nature of $\alpha$-Sn have been initiated on the In-rich InSb(001)-c(8$\times$2) reconstruction of InSb(001)\cite{Anh2021,Barbedienne1954,Barfuss2013,Ohtsubo2013, Rogalev2017, Scholz2018, Chen2022,Madarevic2020,Madarevic2022}, depicted in Fig. 1(a). This reconstruction consists of a 0.5 monolayer (ML) of In and a 0.5 ML of Sb, with another 0.25 ML of In on top\cite{Davis1999}. It is possible that the indium incorporating into the $\alpha$-Sn films is sourced from this surface. On the other hand, the c(4$\times$4) reconstruction, depicted in Fig. 1(b), has 1 ML of In on the surface, followed by 1 ML of Sb, and then an additional 0.75 ML of Sb\cite{Jones1998}. The almost-double layer of Sb on the surface likely decreases the amount of indium that is available on the InSb surface for incorporation into the growing $\alpha$-Sn film. Here we find that by growing $\alpha$-Sn on the Sb-rich InSb(001)-c(4$\times$4), we drastically decrease indium incorporation without distorting the band structure of the epitaxial $\alpha$-Sn films. We show that this reduction in In segregation persists through active heating of the substrate during growth. This method then allows an increase in total heat load that can be applied during $\alpha$-Sn growth, potentially demonstrating a route toward higher mobility thin films. The decrease in indium incorporation is found using \textit{in-situ} angle-resolved photoelectron spectroscopy (ARPES), \textit{in-situ} ultraviolet photoelectron spectroscopy (UPS) and \textit{ex-situ} low temperature magnetotransport. This work paves the way for pure $\alpha$-Sn growth where the features of interest are below the Fermi level, such that topological phase identification through techniques such as ARPES can proceed more readily.

\section{Methods}
\subsection{Sample Growth}
The $\alpha$-Sn films studied here were grown using a modified VG V80 MBE growth system. All samples were grown on undoped (001)-oriented InSb substrates (WaferTech Ltd.). The native oxide was removed via atomic hydrogen cleaning using a thermal cracker (MBE Komponenten) resulting in the In-rich c(8$\times$2)/4$\times$2 surface reconstruction, as determined by reflection high energy electron diffraction (RHEED)\cite{Dong}. The Sb-rich samples were prepared with constant monitoring of the reconstruction via RHEED along the [110] while referencing the InSb(001) surface reconstruction phase diagram\cite{Liu1994,Dong}. The substrate temperature was ramped continuously from 373 K through the following transitions to the final annealing temperature. Near 550 K, the c(8$\times$2) surface was dosed with approximately 0.75 ML Sb until the c(8$\times$2) just fades to the p(1$\times$1) reconstruction. At higher temperatures the reconstruction transitions to the c(4$\times$4) and, starting a few degrees below the c(4$\times$4)/a(1$\times$3) transition (approximately 643 K), was exposed to an Sb$_4$ flux continuously. The sample was then annealed under an Sb$_4$ overpressure 40-60 K above this transition point for at least 30 minutes. The sample was then cooled quickly down back through the a(1$\times$3)/c(4$\times$4) transition to achieve the final c(4$\times$4) reconstruction. Both prepared surfaces show streaky RHEED, indicating relatively smooth surfaces (Fig. 1(c), (d)). The morphology of the substrates was confirmed using 
\textit{in-situ} scanning tunneling microscopy (STM) with the freshly hydrogen cleaned surface (Fig. 1(e)) containing a higher terrace density than after the Sb-termination/anneal procedure (Fig. 1(f)). Some samples with In-rich reconstructions were annealed at 623 K for 30 minutes to smooth the surface further. 

A bilayer (BL) of $\alpha$-Sn is defined here as 9.5$\times 10^{14}$ at/cm$^2$, or $\frac{1}{2}$ of the diamond unit cell. Three different film thicknesses were studied: 13 BL, 50 BL, and 400 BL.  The 13 BL films were grown at a rate of 0.25 BL/min at a substrate thermocouple temperature of 296 K under radiative backside heating. The real temperature is above the melting point of Ga (302.9 K). The 50 BL films were grown at a rate of 0.5 BL/min at a substrate thermocouple temperature of 253 K under passive radiative cooling from the liquid nitrogen cryopanel. The real temperature is significantly below the melting point of Ga. The 400 BL film was grown at a rate of 1.25 BL/min in a separate chamber with active liquid nitrogen cooling on the sample. The sample is indium bonded to a tungsten block which is then in direct thermal contact with an \textit{in-vacuo} liquid nitrogen filled stainless steel vessel. The substrate temperature remains around 80 K during growth. Different growth temperatures and growth rates were used to investigate the effect of heat load on the effectiveness of the termination procedure. The growth rates are calibrated via bilayer oscillations in the RHEED intensity during growth and Rutherford backscattering spectrometry on Si reference samples.
 \begin{figure}
\includegraphics{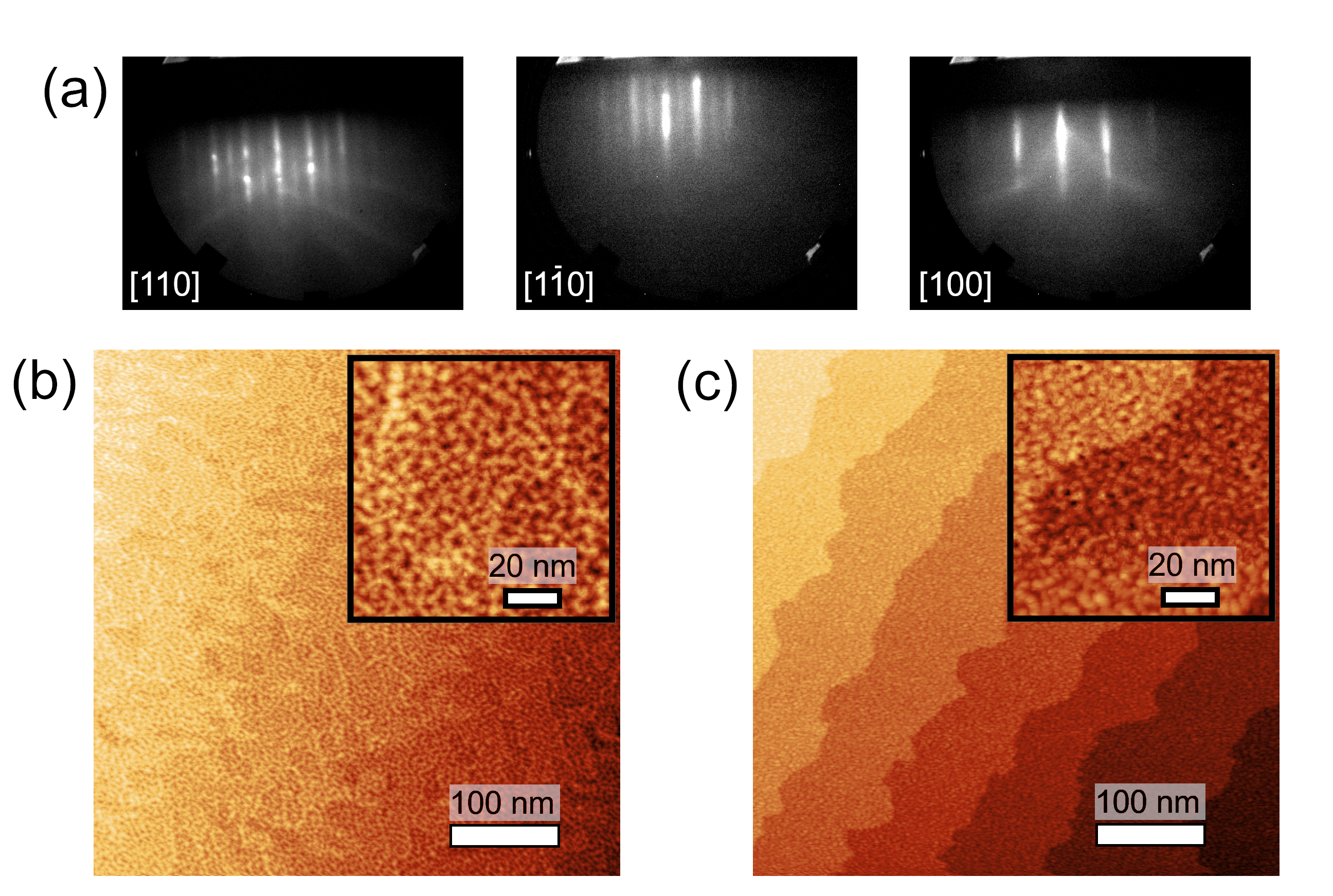}
\caption{\label{Fig2}Surface structure of $\alpha$-Sn(001). RHEED patterns along the [110], [1$\bar{1}$0], and [100] directions showing a 2$\times$, 2$\times$, and 1$\times$ pattern respectively, indicative of a mixed (2$\times$1)/(1$\times$2) reconstruction. \textit{In-situ} STM of 13 BL $\alpha$-Sn grown on (b) InSb(001)-c(8$\times$2) and (c) InSb(001)-c(4$\times$4). Insets are 100 nm x 100 nm measurements under the same conditions.}
\end{figure}
\begin{figure}[!h]
\includegraphics{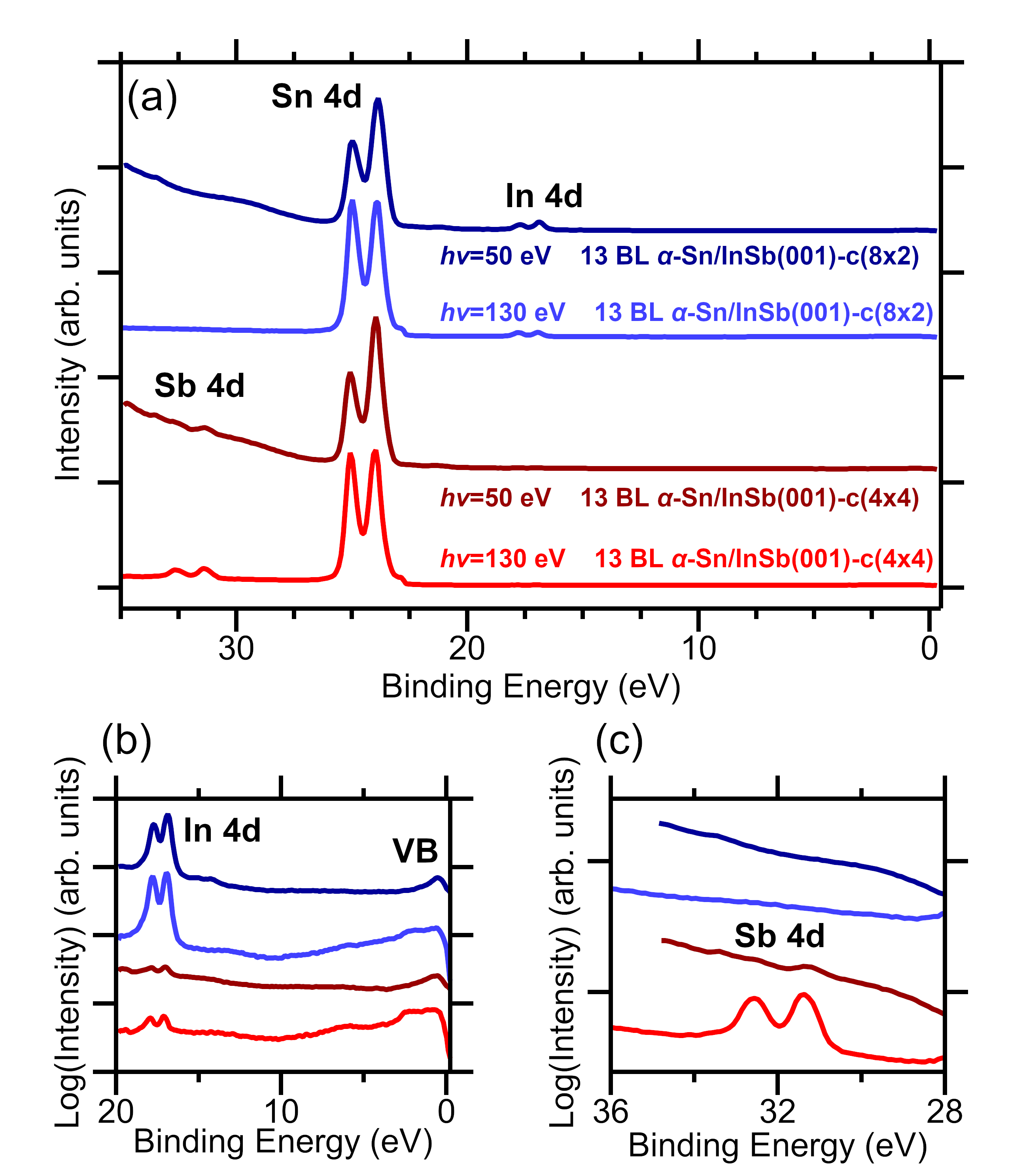}
\caption{\label{Fig3}UPS measurements of 13 BL $\alpha$-Sn films grown on the two different surface reconstructions of InSb(001) with active heating at 296 K. Measurements are offset for clarity. (a) Survey UPS measurements at photon energies of 50 eV and 130 eV, normalized to the integrated Sn 4d intensity. Zoom-in of (b) the valence band and indium core level region and (c) the antimony core level region with the same scan ordering.}
\end{figure}
\subsection{Characterization}
Angle-resolved and -integrated photoelectron spectroscopy measurements of the 13 BL films were performed at beamline 5-2 of the Stanford Synchrotron Radiation Lightsource (SSRL) using linear polarized light. Measurements of the 50 BL and 400 BL film were performed at beamline 10.0.1.2 of the Advanced Light Source (ALS) using \textit{p}-polarized light. Spectra were collected with a Scienta-Omicron DA30L electron analyzer with variable energy resolution and angular resolution better than 0.1$^\circ$. Samples were transferred \textit{in-vacuo} from the MBE system at UCSB to SSRL and ALS via a vacuum suitcase with base pressure better than 4$\times 10^{-11}$ Torr. During the photoemission measurements the sample temperature was kept under 20 K and the pressure was better than 3$\times10^{-11}$ Torr. \textit{In-situ} STM was performed in an Omicron VT-STM at room temperature with a bias of 3 V and a tunneling current of 100 pA. \textit{Ex-situ} magnetotransport measurements on 56 BL films with magnetic field up to 14 T were performed at 2 K in a $^4$He cryostat (Quantum Design Physical Property Measurement System). The 56 BL films were grown under the same conditions as the 50 BL films and were not capped; the native oxide is expected to consume 3-6 BL\cite{Vail2020}. Ohmic contacts were made with silver paint following the van der Pauw geometry. A standard four terminal, low frequency AC lock-in technique was used with a constant excitation current of 100 $\mu$A. 

\section{Results and Discussion}
\subsection{Film morphology}
We observe the typical mixed (2$\times$1)/(1$\times$2) reconstruction in $\alpha$-Sn \cite{Betti2002} characterized by a 2$\times$ RHEED pattern in the $<110>$ directions and a 1$\times$ pattern in the $<100>$ directions (Fig. 2(a)) for all growths less than 100 BL, irrespective of substrate surface termination. During growth all samples show oscillations in the RHEED intensity indicating a bilayer-by-bilayer growth mode for $\alpha$-Sn (see the supplementary material\cite{Supp}). The overall surface morphology is roughly equivalent for both films (Fig. 2(b,c)). The structure consists of bilayer terraces which are then decorated with a grain structure as has been seen previously in $\alpha$-Sn/InSb(001)\cite{Madarevic2020}. The change in bilayer terrace structure between the two cases is likely due to the differences in the starting substrate morphology. The grains in $\alpha$-Sn grown on the Sb-rich surface reconstruction were slightly ($\sim$20\%) smaller than the grains grown on the In-rich surface reconstruction (Fig. 2(b),(c) insets). The RMS roughness on a terrace also decreases from 2.04 $\textrm{\AA}$ to 1.75 $\textrm{\AA}$. The step heights within the grains generally correspond to one atomic layer. The grains are then likely stacks of atomic layers with an alternating (2$\times$1) vs. (1$\times$2) reconstruction, where each layer is partially visible. This then leads to the observed mixed (2$\times$1)/(1$\times$2) reconstruction in RHEED. The alternate stacking structure has been directly observed before in $\alpha$-Sn via STM \cite{KAKU2022151347}, and is well-known for Si(001) surfaces\cite{Tromp}.
\begin{figure}[!h]
\includegraphics{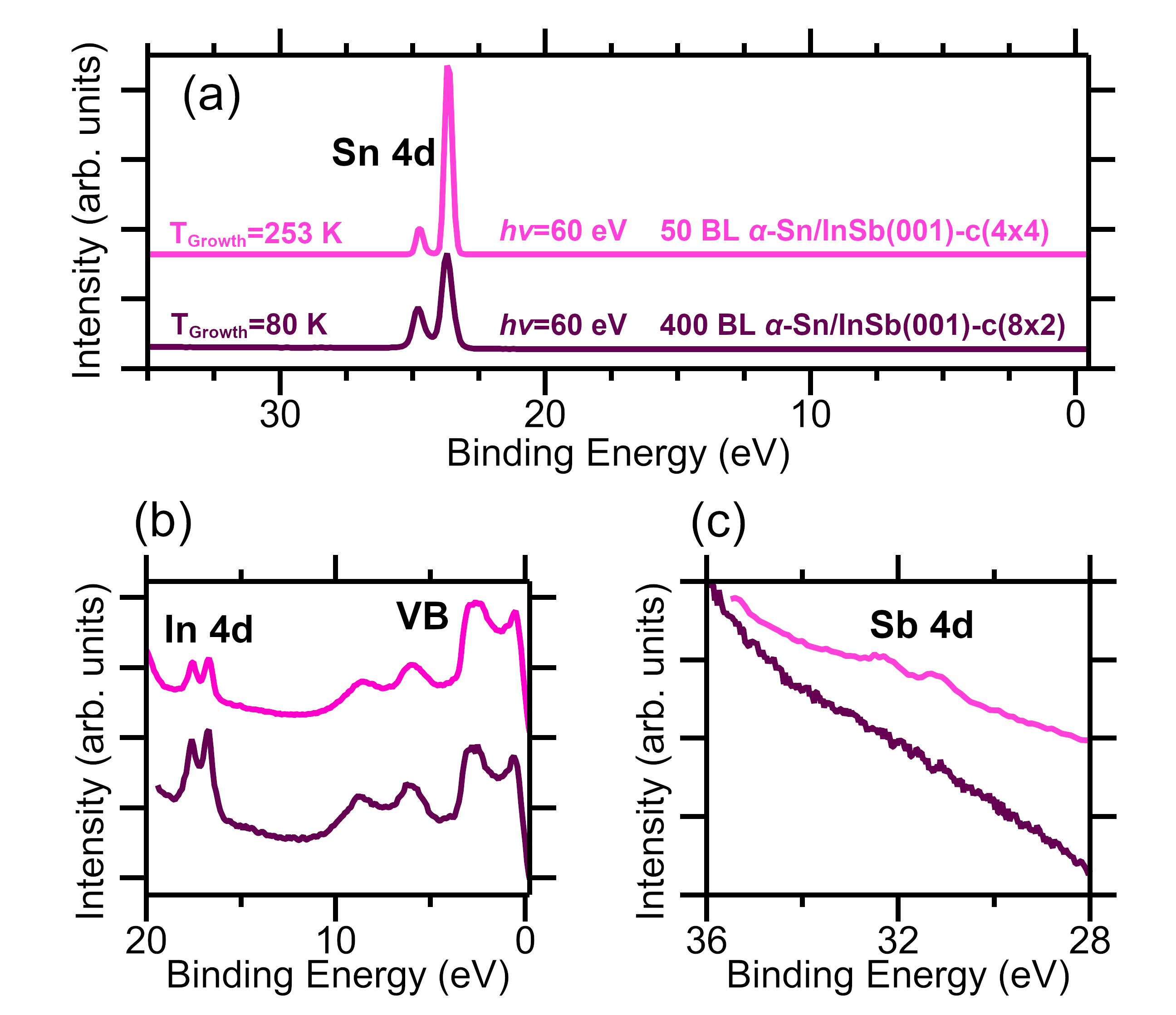}
\caption{\label{Fig4}UPS measurements of 50 BL $\alpha$-Sn on the c(4$\times$4) reconstruction grown at 253 K and 400 BL $\alpha$-Sn on the c(8$\times$2) reconstruction grown at 80 K. Measurements are offset for clarity. (a) Survey UPS measurements at a photon energy of 60 eV, normalized to the integrated Sn 4d intensity. Zoom-in of (b) the valence band and indium core level range and (c) the antimony core level range with the same sample ordering.}
\end{figure}
\begin{figure*}[!ht]
\includegraphics{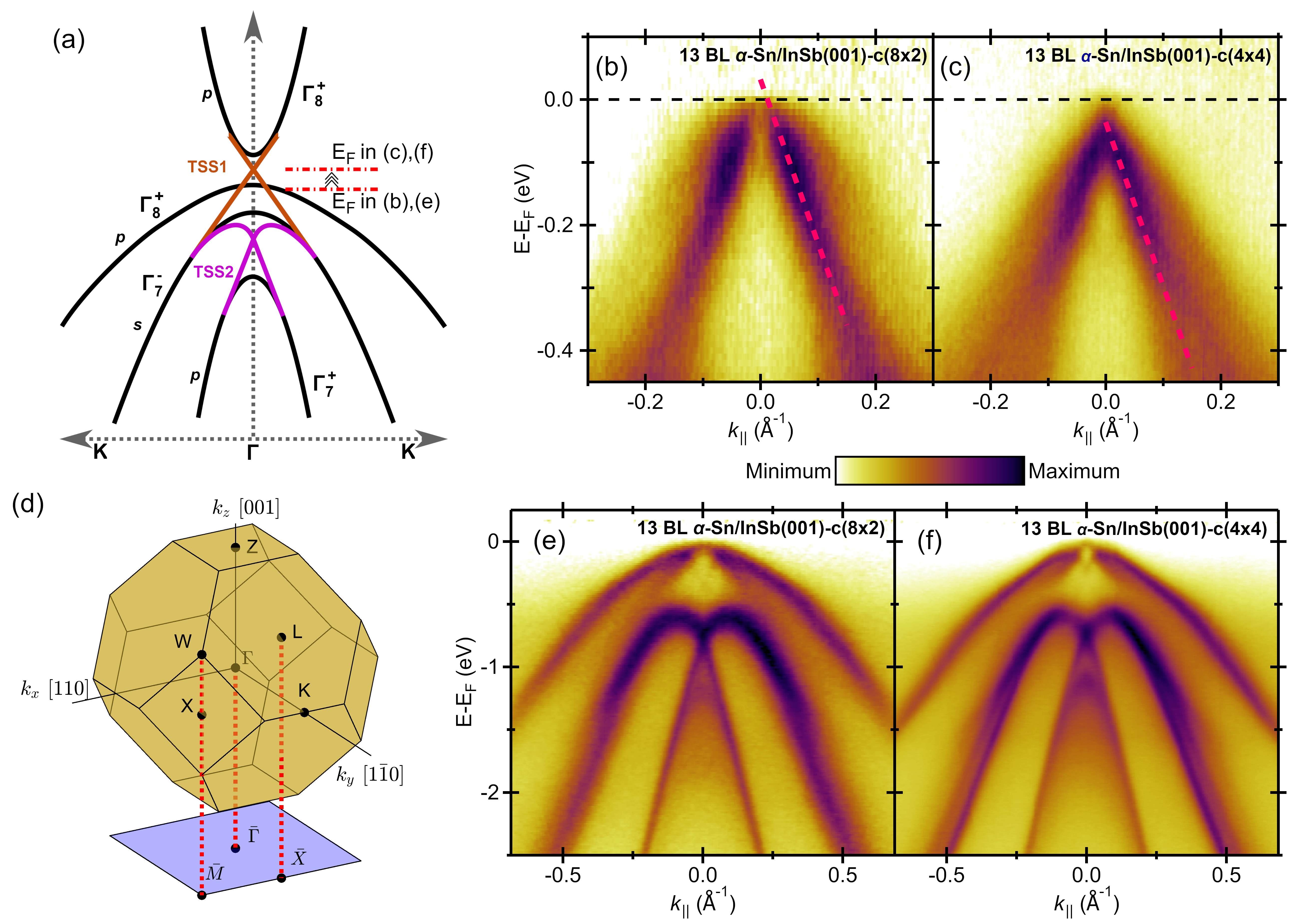}
\caption{\label{Fig5}ARPES measurements of the 13 BL $\alpha$-Sn films. (a) Band schematic of $\alpha$-Sn in a 3D topological insulator-like phase. Surface state measurements at h$\nu$=21 eV in the \bm{$\bar{\textbf{X}}-\bar{\Gamma}-\bar{\textbf{X}}$} direction for 13 BL $\alpha$-Sn grown on (b) InSb(001)-c(8$\times$2) and (c) InSb(001)-c(4$\times$4). Dashed lines corresponding to a linear fit of TSS1 are overlaid on the spectra. (d) Bulk and surface Brillouin zone schematic of $\alpha$-Sn. Bulk band measurements at h$\nu$=127 eV in the \bm{$\textbf{K}-\Gamma-\textbf{K}$} direction in the vicinity of \bm{$\Gamma_{003}$} for $\alpha$-Sn grown on (e) InSb(001)-c(8$\times$2) and (f) InSb(001)-c(4$\times$4).}
\end{figure*}
\subsection{Ultraviolet Photoelectron Spectroscopy}
The effect of substrate termination was first studied on the 13 BL films grown with active substrate heating at 296 K. Measurements were performed at photon energies of 50 eV and 130 eV to deconvolute inelastic mean free path effects, discussed in the supplementary material\cite{Supp}. Growth on the Sb-rich surface termination of InSb(001) resulted in no clear peaks corresponding to indium in a full-range UPS scan (Fig. 3(a)), indicating a large decrease in indium composition. We extracted the very weak intensity of the In 4d states and found a reduction in the indium concentration by over an order of magnitude. Extraction of exact concentration of indium and antimony in the films is problematic as the distribution of dopants is unknown and computed cross-sections and mean free paths under these experimental conditions are inaccurate. Assuming homogeneous composition, using the spectra taken at a photon energy of 130 eV, and using parameters computed by SESSA\cite{SESSA}, we find the total concentration of indium in the film decreases from approximately 2\% to approximately 0.18\% by changing the surface termination that $\alpha$-Sn growth is initiated on. By increasing the signal amplification in the electron analyzer, the indium core levels could be resolved more clearly. With reference to the valence band intensity in Fig. 3(b), a similar sized reduction in the concentration of indium in the films is found. However, there is a concomitant increase in Sb concentration in the $\alpha$-Sn films from growth on the c(4$\times$4) reconstruction (Fig. 3(c)). This is not necessarily a detraction; Sb is an established \textit{n}-type dopant in $\alpha$-Sn\cite{Ewald1968} which should further shift the Fermi level in the desired direction. 

Next we investigated behavior of In and Sb incorporation when $\alpha$-Sn films are grown thicker and with reduced heating during growth. Neither In nor Sb 4d peaks are visible in the survey UPS measurements in Fig. 4(a) for a 50 BL (0.5 BL/min) sample grown at 253 K on an Sb-rich surface nor for a 400 BL (1.25 BL/min) sample grown at 80 K on an In-rich surface. By the same method as earlier we enhanced the signal from the indium, as seen in Fig. 4(b). It is clear that the concentration of indium is reduced further by growing on an Sb-rich surface than by growing at ultracold temperature on an In-rich surface, resulting in a more than 50\% decrease in indium incorporation. There is again an increase in the amount of Sb incorporation (Fig. 4(c)). Thus the concentration of indium in $\alpha$-Sn can be reduced significantly at a range of growth temperatures and growth rates by growing on the Sb-rich c(4$\times$4) reconstruction of InSb(001). This surface termination procedure is more effective than the conventional method of minimizing sample temperature and maximizing Sn growth rate.

\subsection{Electronic Structure}
The persistent \textit{p}-type indium doping in the $\alpha$-Sn films results in the node of the surface states (and the valence band maximum) being above E$_\textrm{F}$ (Fig. 5(a)), and thus not accessible by a filled state measurement such as conventional photoelectron spectroscopy.  This is evident in growth of $\alpha$-Sn on the In-rich surface reconstruction in Fig. 5(b), where the Dirac node is projected to be 32$\pm$7 meV above E$_\textrm{F}$ via a linear fit to the surface states. Reducing the indium incorporation (with the concomitant increase in Sb incorporation) as observed in these same samples in Fig. 3(a), resulted in a Dirac node measured 36$\pm$10 meV below E$_\textrm{F}$ (Fig. 5(c)). The group velocity of the surface states ($\frac{1}{\hbar}\frac{dE}{dk}$ using the slope of the linear fit) in Fig. 5(b) is 4.1$\pm$0.1$\times 10^5$ m/s and in Fig. 5(c) is 4.0$\pm$0.2$\times 10^5$ m/s. The dopants/surfactants of Bi and Te commonly used in $\alpha$-Sn growth result in a high group velocity near 7$\times 10^5$ m/s, while a lower group velocity is found in ``phase pure" $\alpha$-Sn (5$\times 10^5$ m/s)\cite{Madarevic2020,Ohtsubo2013,Scholz2018}. The low and unchanging band velocity we observe is thus consistent with the preservation of “phase pure” $\alpha$-Sn even with the higher concentrations of Sb in the $\alpha$-Sn film. We next turned to the bulk band structure to validate the absence of any band distortion.

The band structure in the vicinity of the \bm{$\Gamma_{003}$} high symmetry point was investigated using a photon energy of 127 eV, assuming an inner potential of 5.8 eV\cite{Rogalev2017}. A schematic of the expected band structure in the case of a 3D topological insulator-like phase is shown in Fig. 5(a)\cite{Rogalev2017, Scholz2018}. In growth on both the In-rich (Fig. 5(e)) and Sb-rich (Fig. 5(f)) reconstruction, the experimental data is consistent with the schematic. There is no evidence of the inverted \textit{p}-like ``light hole"-character conduction band ($\Gamma_8^+$) coming down to touch the VBM, therefore a bulk band gap exists. The topological surface state studied in prior works\cite{Huang2017,Kufner2014,Barbedienne1954,Barfuss2013,Chen2022,Ohtsubo2013,deCoster2018,Rogalev2017,Rogalev2019,Scholz2018,Xu2017,Madarevic2020} (and which determines the topological phase) is labelled TSS1; its Dirac node is located approximately 80 meV above the valence band maximum. The valence band maximum consists of the uninverted ``heavy hole" band with \textit{p}-character ($\Gamma_8^+$). Below this is the inverted “conduction” band with \textit{s}-like character ($\Gamma_7^-$). The lowest depicted band  is the \textit{p}-like split-off band ($\Gamma_7^+$). The distinctive \textit{M} shaped feature between the split-off band and the “conduction” band are associated with a second topological surface state TSS2, discussed in more detail elsewhere\cite{Rogalev2017}. No discernible change to the bulk dispersion is observed, indicating the reduction of indium and addition of antimony can be treated as a rigid band shift from doping. While carrier density changes could be extracted from the knowledge of \textit{k$_\textrm{F}$}, surface band bending is a non-negligible effect in topological insulators\cite{BRAHLEK201554}.  As ARPES (and UPS) are surface sensitive measurements, these observed carrier density changes do not reflect accurate changes in the carrier densities in the bulk of the crystal (i.e. an estimation of the doping). To ensure the reduction in indium incorporation is reflected in the bulk of the film, we turned to low temperature magnetotransport. 
\begin{figure}[h]
\includegraphics{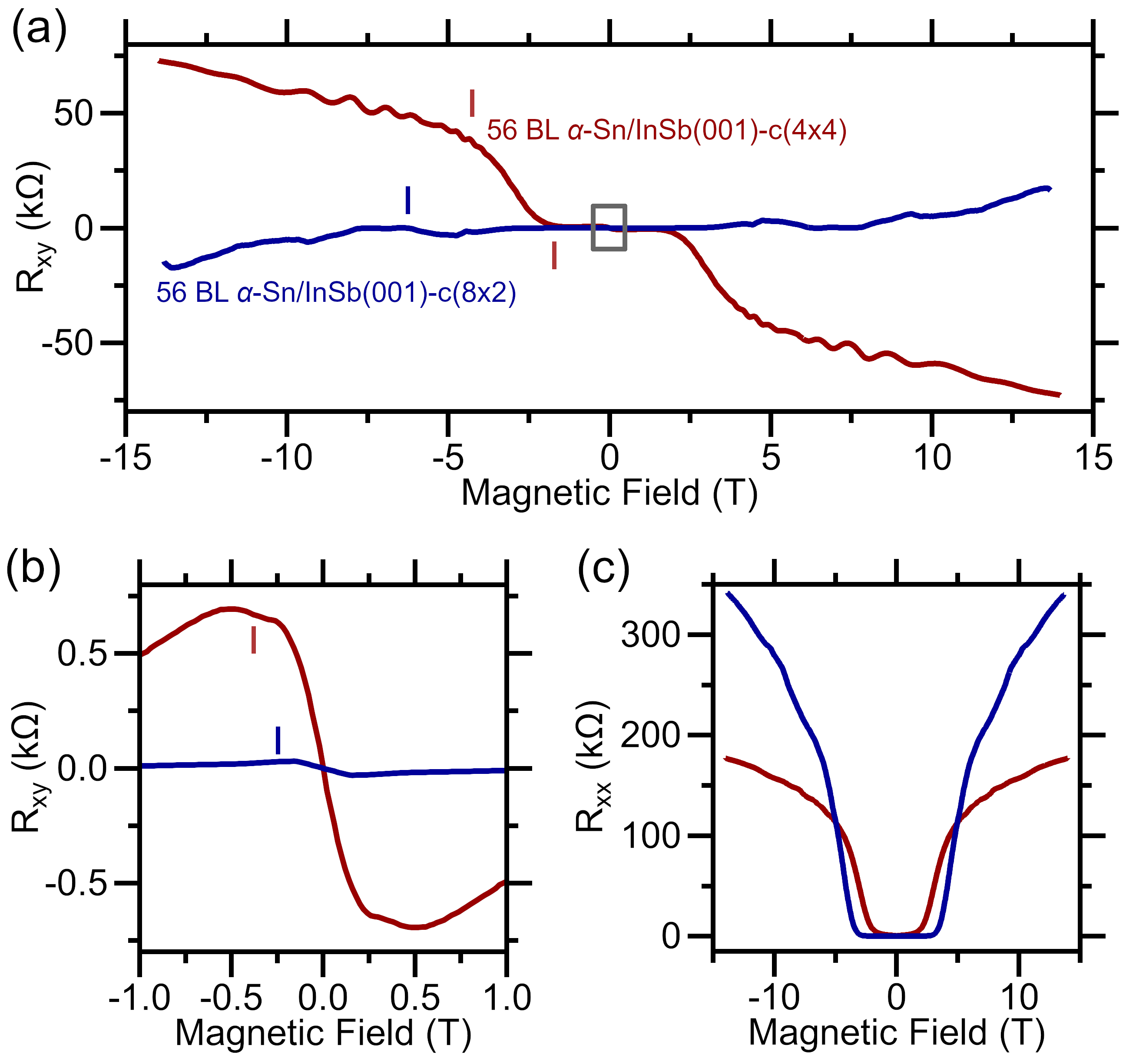}
\caption{\label{Fig6} Magnetotransport measurements of 56 BL $\alpha$-Sn films performed at 2 K. (a) Transverse resistance as a function of magnetic field for $\alpha$-Sn films grown on the c(8$\times$2) and c(4$\times$4) surfaces of InSb(001). (b) Zoom-in of (a) showing an additional inflection point in each sample. (c) Longitudinal resistance vs. magnetic field for the two samples. Approximate inflection points in transverse geometry measurements are marked.}
\end{figure}
\subsection{Magnetotransport}
Longitudinal and transverse resistance measurements for 56 BL $\alpha$-Sn films initiated on the two different InSb(001) reconstructions are shown in Fig. 6. Growth on InSb(001)-c(4$\times$4) resulted in \textit{n}-type behavior at high field while growth on InSb(001)-c(8$\times$2) resulted in \textit{p}-type behavior at high field in Fig. 6(a). Shubnikov-de Haas oscillations are visible in both the longitudinal and transverse geometry. Additional low field Hall measurements in Fig. 6(b) clarify the presence of three inflection points in the Hall effect in $\alpha$-Sn on Sb-rich InSb and only two inflection points in $\alpha$-Sn on In-rich InSb. This change is likely due to an increase in mobility of the bulk $alpha$-Sn carriers. The bulk mobility is sensitive to crystal quality, the carriers only being observable in quantum oscillations in high quality samples\cite{Madarevic2022,Barbedienne1954, Anh2021}. The onset of Shubnikov-de Haas oscillations occurs at a lower magnetic field value in films grown on Sb-rich InSb(001), indicating higher quantum mobilities\cite{Shoenberg1984}. We anticipate this initial onset to be indicative of the surface states, which have been found to have a higher mobility than the bulk heavy holes\cite{Madarevic2022,Barbedienne1954,Anh2021}. In the longitudinal geometry, we observe a sharp increase in the resistance at an applied field of around 4 T (Fig. 6(c)). This is a consequence of the established magnetic field induced metal-insulator transition in InSb\cite{Ishida2003}. The behavior of freeze-out in InSb is extremely sensitive to absolute dopant concentrations\cite{vonOrtenberg}. Carrier freeze-out complicates Hall analysis, as the response of the substrate in a transverse geometry is highly non-linear and has no simple analytical form. While quantitative analysis is difficult, we find that there is a clear, drastic change in the carrier concentrations in the $\alpha$-Sn films when grown on different surface terminations of InSb(001). The magnetotransport concurs with both the UPS and ARPES measurements that growth on the Sb-rich InSb(001) surface reconstruction reduced \textit{p}-type doping in $\alpha$-Sn films by limiting indium incorporation.

\section{conclusion}
Topologically non-trivial $\alpha$-Sn thin films have been grown on the Sb-rich InSb(001)-c(4$\times$4) and In-rich InSb(001)-c(8$\times$2) surface reconstructions. Despite active substrate heating and lack of intentional surfactant or dopant species, the $\alpha$-Sn films grown on the Sb-rich reconstruction show minimal indium incorporation. This method results in less indium incorporation than even growing $\alpha$-Sn at cryogenic temperatures while on the In-rich c(8$\times$2) reconstruction. The reduction in \textit{p}-type doping was confirmed through UPS, ARPES, and magnetotransport. Our work facilitates more robust identification of the topological phase via angle-resolved photoemission and magnetotransport. Furthermore, it allows for a wider growth window of $\alpha$-Sn thin films on InSb(001) and opens up a path for a similar methodology on other crystal orientations of interest in this materials system. 

\begin{acknowledgments}
The growth, magnetotransport and later ARPES studies were supported by the Army Research Laboratory (W911NF-21-2-0140 and W911NF-23-2-0031). The initial vacuum suitcase construction and initial ARPES measurements were supported by the US Department of Energy (DE-SC0014388). The UC Santa Barbara NSF Quantum Foundry funded via the Q-AMASE-i program under award DMR-1906325 support was used for further development of the vacuum suitcases. This research used resources of the Advanced Light Source, which is a DOE
Office of Science User Facility under contract no. DE-AC02-05CH11231.  Use of the Stanford Synchrotron Radiation Lightsource, SLAC National Accelerator Laboratory, is supported by the U.S. DOE, Office of Science, Office of Basic Energy Sciences under Contract No. DE-AC02-76SF00515. The authors would like to further thank A. M. Kiefer, G. J. de Coster, P. J. Taylor, P. A. Folkes, and O. A. Vail for fruitful discussions.
\end{acknowledgments}


\bibliography{export}


\end{document}